# Angular momentum distributions for observed and modeled exoplanetary systems

Jonathan H. Jiang[1], Remo Burn[2], Kristen A. Fahy[1], Xuan Ji[3], Patrick Eggenberger[4]

[1]Jet propulsion Laboratory, California Institute of Technology, Pasadena, California, U.S.A
[2]Max-Planck Institute for Astronomy, Heidelberg, Germany
[3]Department of The Geophysical Sciences, University of Chicago, Chicago, Illinois, USA
[4]Département d'Astronomie, Université de Genève, Genève, Switzerland

Corresponding to: Jonathan.H.Jiang@jpl.nasa.gov

**Abstract**

The distribution of angular momentum of planets and their host stars provides important information on the formation and evolution of the planetary system. However, mysteries still remain, partly due to bias and uncertainty of the current observational datasets and partly due to the fact that theoretical models for the formation and evolution of planetary systems are still underdeveloped. In this study, we calculate the spin angular momenta of host stars and the orbital angular momenta of their planets using data from the NASA Exoplanet Archive, together with detailed analysis of observation dependent biases and uncertainty ranges. We also analyze the angular momenta of the planetary system as a function of star age to understand their variation in different evolutionary stages. In addition, we use a population of planets from theoretical model simulations to reexamine the observed patterns and compare the simulated population with the observed samples to assess variations and differences. We found the majority of exoplanets discovered thus far do not have the angular momentum distribution similar to the planets in our Solar System, though this could be due to the observation bias. When filtered by the observational biases, the model simulated angular momentum distributions are comparable to the observed pattern in general. However, the differences between the observation and model simulation in the parameter (angular momentum) space provide more rigorous constraints and insights on the issues that needed future improvement.

## 1. Introduction

Technological and scientific advancements have rapidly expanded the field of exoplanets and have allowed us the incredible opportunity to study and compare the over 3,200 star-systems discovered to date. Using this observational data, we have been able to deduce that the Solar System is indeed unique in many ways. For example, the orbits of our planets are nearly circular, and unlike the observations of many systems, our four giant planets are at a much further distance than the Sun, with smaller planets existing closer- a cosmic pattern that appears uncommon in comparison to the exoplanetary systems who host giants much closer with highly elliptical orbits [*Beer et al.* 2004].

It is therefore important to examine how our Solar System, and other star systems, vary with respect to angular momenta and look at the distributions between observations versus models for a variety of systems. For this purpose, a combined observational and modeling analysis to compare exoplanetary systems for both the total angular momentum as well as the total orbital angular momentum of the planets is needed.

The distribution of angular momentum between host star and planets provides essential information regarding the formation and evolution of the planetary system. Many planetary formation theories, before the discovery of exoplanets, were based upon our own Solar System objects. Early studies by *Fish* [1967] examined angular momentum densities for the planets and planet-satellite systems including asteroids and the Earth-Moon system. Since then, many studies have been conducted to further our understanding of angular momentum in exoplanetary systems, though it remains an area in which we lack insight on the full assessment between theory and observation. A major constraint with any exoplanetary study is the quality and amount of data necessary for meaningful large-scale analysis. Now with the increase available exoplanet data, we can make comparisons on a much broader scale than ever before with hundreds of datapoints to elucidate patterns. However, the observational biases related to the different detection techniques (e.g., radial velocity (RV) and transit method, etc.) could mislead us such that any patterns we find might not be linked to the physical meaning we seek.

On the other hand, models for planetary angular momentum have been created to study the orbital mechanics of exoplanets such as *Gurumath et al.* [2018], who were able to use real data to estimate orbital angular momentum of known exoplanets. Models are also developed to study the angular momentum for specific ranges of host stars. For example, a theoretical model by *Stills et al.* [2002] focused on low mass stars (0.1-0.5 $M_\odot$) and solar analogs (0.6-1.1 $M_\odot$); *Gallet & Bouvier* [2013] presented specific models for the "rotational evolution of solar-like stars between 1 Myr and 10 Gyr with the aim of reproducing distributions of rotational periods observed for star forming regions and young open clusters within this age range"; furthermore, *Matt. et. al.* [2015] sought to understand the "observed distributions of rotation rate and magnetic activity of sunlike and low-mass stars".

While model function and sensitivity have improved over time as we see with *Gallet & Bouvier* [2015] who added in a wind braking law based on magnetized stellar winds and found sensitive dependencies in their model related to mass and instantaneous rotation rate, there still remains an inherent lack of comprehensivity.

In summary, the main gaps that exist in our present understanding of angular momentum in exoplanetary systems are due to a) bias and uncertainty of the current observational datasets, and b) theoretical models for the formation and evolution of planetary systems are presently underdeveloped.

This study aims to provide a quantitative assessment of spin angular momenta of host stars and the orbital angular momenta of their planets using observational data. In order to understand variation of angular momentum in different evolutionary stages of stars, we bin the observational data according to the known age of the stars and use a model simulation to analyze changes in angular momenta of planetary systems over their lifetime. The angular momenta generated by the model simulation are filtered by the observational biases, in order to re-examine the correlations and compare the simulated population with the observed samples. In doing so, we can examine the crucial topic of planetary formulation. Do the exoplanetary systems discovered thus far have an angular momentum distribution similar to our Solar System planets? We hope our analyses will bring us closer to determining just how special or perhaps ordinary, we really are.

The organization of the paper is as follows: Section 2 describes the methodologies for calculating the angular momenta from the observational data, estimating the observational biases, model simulations and the corresponding parameters to compare with the observations. Subsequently, section 3 analyzes (a.) the distribution of angular momenta of the exoplanet systems from observed data and compares them with our Solar System; (b.) the comparison between observation and model simulations at different time stages of star and planetary systems.

## 2. Methodology
### 2.1 Calculating the angular momenta from observational data

The angular momenta in a planetary system includes both the orbital angular momenta of the planets and the rotational angular momentum of the star (here we neglect the spin angular momentum of the planets). These can be computed from the systems when the mass and rotation of the host star, and the mass, semi-major axis, eccentricity and inclination of the planets are known.

For multi-planet systems, we consider the component of orbital angular momentum which is parallel to the stellar spin axis.

The total orbital angular momenta of all the planets in a system are:

$$L_p = \sum_{n=1}^{n=N} \sqrt{GM_* m_n^2 a_n (1-e_n^2)} \cos(i_n) \qquad (1)$$

where $G$ is the gravitational constant, $M_*$ is the mass of the star, $N$ is the total number of planets in the system, subscript $n$ denotes the $n^{\text{th}}$ planet in the system with mass $m_n$, semi-major axis $a_n$, eccentricity $e_n$, and inclination $i_n$.

The spin angular momentum of the host star, $L_*$, is estimated with the following equation following *Gurumath et al.* [2019] which only takes radial rotational velocity into account:

$$L_* = \frac{v \cdot \sin i}{R_*} I_* \qquad (2)$$

where $i$ is the inclination about the rotation axis, $v$ is the rotational velocity of the star, $R_*$ is the radius of the star, and then

$$I_* = \frac{2}{5} M_* R_*^2$$

is the moment of inertia of the star calculated with a reasonable assumption of rigid body. So, equation (2) can be rewritten as:

$$L_* = \frac{2}{5} v \cdot \sin i \cdot M_* R_* \qquad (3)$$

Let $\sigma_x$ be the measurement uncertainty of a parameter $x$ listed in the NASA Exoplanet Archive. According to the propagation of uncertainties, the relative uncertainty of orbital angular momentum of planet $n$ is:

$$\sigma_{L_n} = L_n \sqrt{\left(\frac{\sigma_{M_*}}{2M_*}\right)^2 + \left(\frac{\sigma_{m_n}}{m_n}\right)^2 + \left(\frac{\sigma_{a_n}}{2a_n}\right)^2 + \left(\frac{-e_n \sigma_{e_n}}{1-e_n^2}\right)^2} \qquad (4)$$

The uncertainty of the total angular momentum of the planetary system is:

$$\sigma_{L_p} = \sqrt{\sum_n \sigma_{L_n}^2} \qquad (5)$$

The uncertainty of the spin angular momentum of the host star is:

$$\sigma_{L_*} = L_* \sqrt{\left(\frac{\sigma_{v \cdot \sin i}}{v \cdot \sin i}\right)^2 + \left(\frac{\sigma_{M_*}}{M_*}\right)^2 + \left(\frac{\sigma_{R_*}}{R_*}\right)^2} \qquad (6)$$

To ensure good quality of observational data used in this study, we only accept data satisfying $\sigma_{L_p}/L_p < 1$ and $\sigma_{L_*}/L_* < 1$.

## 2.2 Observational bias

The transit method is highly biased towards planets on very short period orbits (~ days) and closer to the host star (< 1 astronomical unit, AU), because they transit more frequently [e.g., *Jiang et al.* 2019]. The transit method is also biased towards discovering large planets, because larger planets block more lights and thus are easier to detect, or biased towards finding big planets around small stars for the same reason. The RV method is biased towards finding massive exoplanets since the doppler effect is more noticeable with large planets orbiting close to its host star [e.g., *Zakamska et al.* 2011]. Microlensing and direct imaging methods are most sensitive to detecting planets that are more than 1 AU away from Sun-like stars [e.g., *Tsapras* 2018]. Another consideration is that the total angular momentum in a planetary system will vary due to the number of planets in the system. There might be planets undiscovered in the systems especially those further away from the host, likely being the case for Kepler and TESS. Furthermore, the observational datasets are also highly heterogeneous in terms of relative precision on the measurements and types of host stars.

Due to limitations of the observation methods, the angular momenta are mostly calculated using data from two observational methods, transit and RV, each having different pros and cons. The exoplanet radius is calculated by the depth of transit, and the exoplanet mass is calculated using RV based measurements of velocity along line of sight. In the NASA Exoplanet Archives, almost all the planet radius data are discovered via transit method, and planet mass is often obtained first by the transit discovery of planet, e.g., with Transit Timing Variation (TTV) or using a mass-radius relationship [*Agol et al.* 2005; *Holman* 2005], and then subsequently estimated with follow-up RV observations.

Most of the exoplanets in our angular momenta analyses were discovered via the transit method, specifically by the Kepler telescope, and it requires more than 2 transits to confirm a planet, which limited the orbital period of detectable planets. As Kepler worked for 4 years, the planets with orbital period larger than ~730 days cannot be detected by Kepler, which further limited the detectable angular momenta.

The masses of these planets were later determined by RV after the transit discovery. The amplitude $K_1$ for RV detection is in terms of eccentricity ($e$), planet mass ($M_p$), stellar mass ($M_*$), and orbital period ($P$) is expressed as [*Lovis & Fischer* 2010]:

$$K_1 = \frac{28.4329 ms^{-1}}{\sqrt{1-e^2}} \frac{M_p \sin i}{M_J} \left(\frac{M_* + M_p}{M_\odot}\right)^{\frac{2}{3}} \left(\frac{P}{1yr}\right)^{-\frac{1}{3}} \quad (7)$$

where $M_J$ is mass of Jupiter and $M_\odot$ is solar mass.

Our analysis found that almost all of the scaled amplitudes of the detected planets are $K_1 > 10$ m/s. We therefore adopt this as the detection threshold. Assuming the inclination is close to 90 degrees and eccentricity is close to zero, $K_1$ can be written in terms of

$$K_1 \propto \frac{M_p^2 M_*^{4/3}}{L} \left(1 + \frac{M_p}{M_*}\right)^{2/3} \tag{8}$$

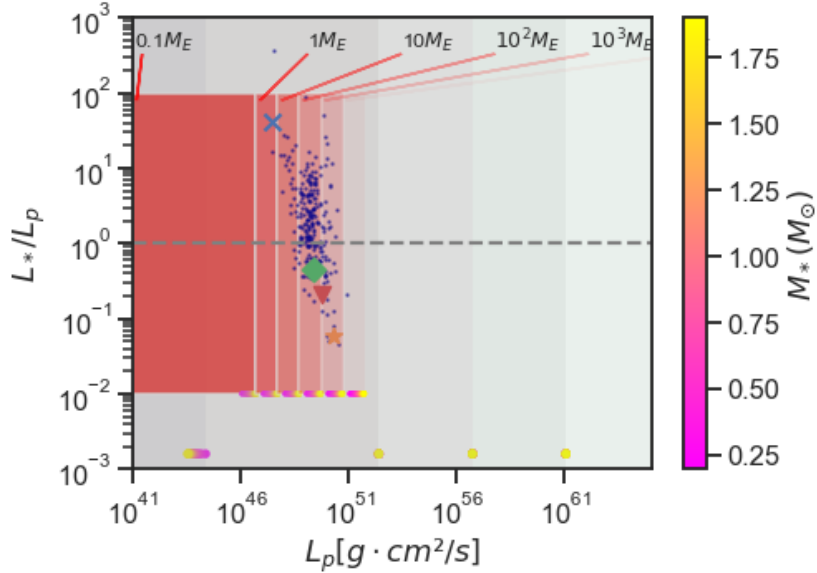

**Figure 1:** Scatter plot of planet orbital angular momenta versus the ratio of stellar rotational angular momenta to planet orbital angular momenta. The background color blocks indicate the upper limit of orbital angular momenta of the planets detectable by RV, with masses that are shown at the top of the figure, e.g. a planet with $10M_\oplus$ can only be detected by the RV method if its orbital angular momentum is $\lesssim 10^{57} g \cdot cm^2/s$. The horizontal color sticks near the bottom of the figure also mark the upper limit of planet angular momenta detectable by RV, with mass consistent with the color blocks and with the host star mass suggested by the color-bar. At the left side of the figure, the smaller foreground color marks the upper limit of orbital angular momenta of planets that detectable by the transit method, with given masses as pointed to by the arrows. The horizontal color sticks at the bottom of the smaller foreground blocks show the transit detectable limits, with masses that are consistent with the color blocks and with host star mass suggested by the color-bar. The blue cross, orange star, green diamond, and red triangle mark Earth, Jupiter, Neptune and Jupiter at Mercury's orbit, respectively.

This shows that for a given star, the RV amplitude limits the possible angular momentum of a detectable planet with mass $M_p$. To simplify this problem, we look at the observational bias for single-planet systems and focus on potential bias in orbital angular momentum distribution. We assume that the stellar mass ranges from $0.2$-$2M_\odot$, planetary mass ranges from $0.1$-$10^4 M_\oplus$. We then compute the limits caused by the detection methods of RV and transit separately, which are shown in Figure 1. It suggests that only if the planetary mass is $< 1M_\oplus$, will it suffer the RV bias.

Figure 2 shows the histogram of $L_p$ of the planets with only $M_p\sin i$ information, which can be thought of as those without transit follow-up observation. We still found the drop, e.g., at $\sim 10^{52}$ $g \cdot cm^2/s$, although there are more planets with higher $L_p$. As $M_p\sin i$ is the minimum mass, the true $L_p$ could be higher. Whether or not it is a physical drop, however, is unclear, as we already know their masses and most of them are beyond $1M_\oplus$. We would argue it is because almost all of the RV detected planets have an orbital period $< 10^4$ days, which is 28 years (which is equivalent to $\sim 8$ AU when orbiting the Sun), and a jupiter-mass planet at 10 AU is $2\times 10^{50}\, g \cdot cm^2/s$. Therefore, it is limited by the maximum detectable orbital period.

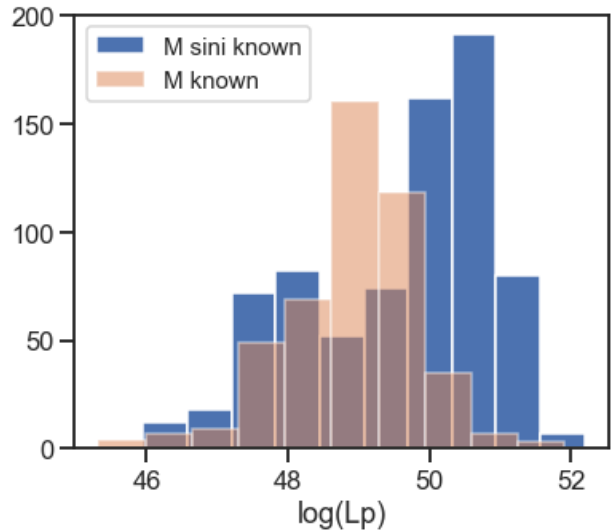

**Figure 2:** Histogram of $L_p$ of the planets with M information (orange) and with only $M_p\sin i$ information (blue), which are those without transiting follow-up observation.

Also shown in Figure 1 is that, at the same time, all of the planets in our sample suffer from transit bias, but at varying levels for different planetary masses. Analyzing the observational biases as discussed above will help us to better understand the current data sample and help infer the true distribution of the physical parameters.

**2.3 Angular momenta from model simulations and filtered by observational bias**

Angular momenta of planetary systems can be simulated by means of a *planetary population synthesis* model. This approach relies on observed distributions of protoplanetary disk properties (i.e., mass, metallicity, inner edge, and lifetime) to then compute a population of synthetic planets. [*Ida & Lin* 2004, *Mordasini et al.* 2009, *Benz et al.* 2014, M*ordasini* 2018]. *Emsenhuber et al.* [2021b] and *Burn et al.* [2021] utilize a recently updated planetary synthesis model [*Emsenhuber et al.* 2021a] to simulate the angular momenta of planetary systems.

Here, we briefly outline a number of key ingredients to this model. The gaseous disk evolves α-viscously [*Shakura & Sunyaev* 1973] and the midplane temperature is analytically approximated [*Nakamoto & Nakagawa* 1994]. It depends on irradiation from the star [*Hueso & Guillot* 2005] and viscous heating as well as the mean opacity [*Bell & Lin* 1994].

The modeled growth of a protoplanet starts at the stage of a lunar-mass embryo, which is assumed to have grown in a stage of runaway planetesimal accretion [*Kokubo & Ida* 1998]. Multiple such protoplanets are injected at random locations in the disk consisting of gas and planetesimals. Then, the accretion rates of planetesimal [*Fortier et al.* 2013] and gas, as well as the gravitational interactions between the multiple growing protoplanets using the MERCURY N-body code [*Chambers* 1999] are calculated. To derive realistic accretion rates of gas, the standard equations governing the evolution of the gaseous envelope [*Bodenheimber & Pollack* 1986] are solved. In addition to that, protoplanet migration in the type I [*Paardekooper et al.* 2010,2011] or type II [*Dittkrist et al.* 2014] regime is considered.

The gaseous disk is observed to disperse quite rapidly [*Haisch et al.* 2001], which is achieved in the models by including internal and external disk photoevaporation [*Clarke et al.* 2001, *Matsuyama et al.* 2003]. The latter is scaled with a stochastic parameter to reproduce the observed lifetime distribution of disks.

After the dissipation of the gas disk, the planets are able to lose their accretional energy over Gyr timescales. For that, the model continues to solve the internal structure of the protoplanets including stellar irradiation and photoevaporation of the atmospheres [*Jin et al.* 2014].

During this late stage of the evolution, the star also significantly evolves. In the model, during all stages, the stellar temperature, luminosity, and radius follows the evolutionary tracks of *Baraffe et al.* [2015]. This is not the case for the rotation period of the star $P_\star$ which is key to this project. We therefore parametrize the evolution of $P_\star$ to match reasonably well with the resulting surface rotation periods of *Eggenberger et al.* [2019] and *Amard et al.* [2019,2020]:

$$P_\star(t) = \begin{cases} P_{\text{ini}} & \text{for } t \leq t_{\text{disk}} \\ P_{\text{ini}} \left(\dfrac{t}{t_{\text{disk}}}\right)^{-0.435} & \text{for } t_{\text{disk}} < t \leq t_{\text{PMS}} \\ \left(\dfrac{1}{P_\star(t_{\text{PMS}})} - \dfrac{a}{P_\odot}\log_{10}^2\left(\dfrac{t}{t_{\text{PMS}}}\right) - \dfrac{b}{P_\odot}\log_{10}\left(\dfrac{t}{t_{\text{PMS}}}\right)\right)^{-1} & \text{for } t_{\text{PMS}} < t \leq t_{\text{conv}} \\ P_{\text{late}} \left(\dfrac{t}{t_{\text{PMS}} - t_{\text{disk}}}\right)^{0.5} 10^{0.236\,[\text{Fe/H}]} & \text{for } t > t_{\text{conv}} \end{cases} \quad (9)$$

For this parametrization, $t_{\text{disk}}$ is the lifetime of the protoplanetary disk, which varies for each simulation, but ultimately samples the distribution of observed disk lifetimes [e.g., *Haisch et al.* 2001,

*Richert et al.* 2018]. $P_{\text{ini}}$ is the initial rotation period chosen to sample the measured distribution by *Venuti et al.* [2017], specifically for young objects with disk signatures in the open cluster NGC 2264. This quantity is also used to describe the inner edge of the disk in the planetary population synthesis model. The parameter $b$ is set to the value 6 and $a$ is chosen to get a continuous evolution function. The rest of the parameter choices for the fit are listed in Table 1 and vary for different stellar masses. In particular, the pre-main sequence time $t_{\text{PMS}}$ is very different for different masses. We note that for late-type M dwarfs ($M_\star < 0.5$), $t_{\text{PMS}}$ should be instead be considered as a time of maximum rotation. For the discussion, we note that late-stage rotation of similar stars is almost indistinguishable despite different initial rotation periods [*Eggenberger et al.* 2019] and results converge towards the classical result of a power-law with a slope of 0.5 [*Skumanich* 1972]. To finally get the angular momentum of the modeled stars, we further assume rigid-body rotation and uniform densities based on the *Baraffe et al.* [2014] radii.

**Table 1:** Stellar-rotation model parameters

|  | $M_\star: 1.0\ M_\odot$ | $M_\star: 0.7\ M_\odot$ | $M_\star: 0.5\ M_\odot$ | $M_\star: 0.3\ M_\odot$ | $M_\star: 0.1\ M_\odot$ |
|---|---|---|---|---|---|
| $t_{\text{PMS}}$ | $4.0 \times 10^7$ yr | $7.9 \times 10^7$ yr | $1.2 \times 10^8$ yr | $3.2 \times 10^8$ yr | $6.3 \times 10^8$ yr |
| $P_{\text{late}}$ | $0.082\ P_\odot$ | $0.0185\ P_\odot$ | $0.294\ P_\odot$ | $0.312\ P_\odot$ | $0.312\ P_\odot$ |
| $t_{\text{conv}}$ | $10^8$ yr | $3.6 \times 10^8$ yr | $8.3 \times 10^8$ yr | $1.3 \times 10^9$ yr | $1.6 \times 10^9$ yr |

The resulting stellar angular momenta are shown in Figure 3 where we used solar densities for simplicity. The evolution compares reasonably well to the detailed models of *Amard et al.* 2019 and to observational data collected by *Gallet & Bouvier* [2015].

Because exoplanets are frequently discovered around stars with sub-solar masses, we include in the synthetic dataset the results of *Burn et al.* [2021] who discuss the synthetic planetary population for K and M dwarfs in addition to the nominal solar-like case. To achieve this, the initial conditions for the protoplanetary disks need to be scaled. The disk mass is assumed to depend linearly on the stellar mass as discussion in *Burn et al.* [2021] and disks around solar mass stars are distributed according to the VLA measurements of Class I disks by *Tychoniec et al.* [2018]. These young objects are found to be surrounded by disks that are an order of magnitude more massive (median dust mass is 96 $M_\oplus$) than the older, more common Class II objects ($\sim 10\ M_\oplus$). We note, that more recent works find lower disk masses for the same class of objects [i.e., *Williams et al.* 2019]. However, there is the additional constraint of measured gas accretion rates onto the stars [e.g., *Hartmann et al.* 2018], which are favoring larger disks.

In addition to initial masses, dust-to-gas ratios are distributed equally for all stellar masses following the measurements of *Santos et al.* 2003. Initial locations of protoplanets are drawn from a log-uniform distribution of semi-major axis spanning from the inner edge to separations corresponding to orbital periods of 253.2 yr. Model parameters that are fixed for all simulation runs can be found in Table 2. Addionally, references and further discussion can be found in *Burn et al.* [2021]).

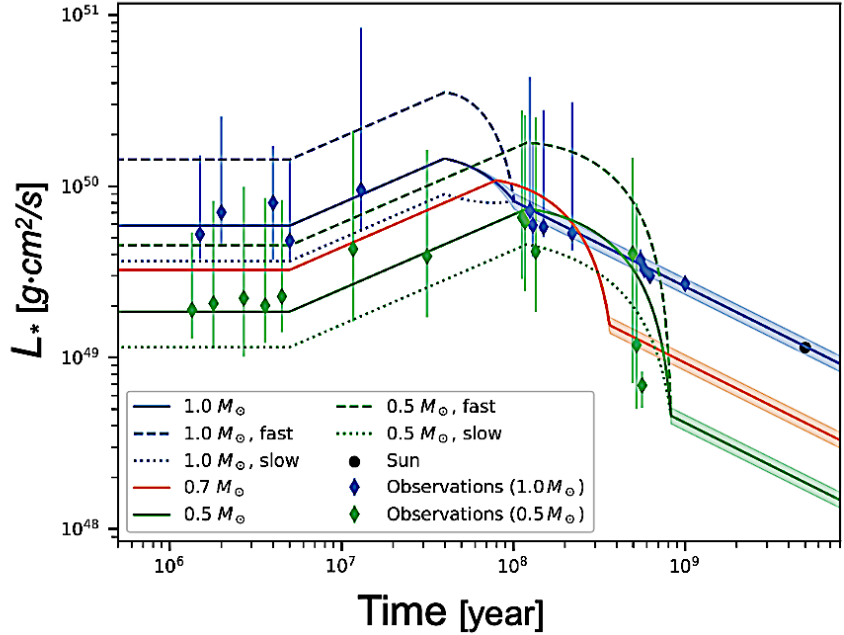

**Figure 3:** Parametrized angular momentum evolution for different stellar masses and initial rotations. Observations of angular momenta of different star forming regions and open clusters compiled by *Gallet & Bouvier* [2015] are shown for comparison. To transform rotation rates to angular momenta, we assumed solar densities and rigid body rotation. Observational data for 0.5 $M_\odot$ are shifted back in time by 10 % for better visibility. The error-bars, respectively "fast" and "slow" cases, denote the 25th to 90th percentile range. The simple fit does not capture all transition phases accurately, but a good agreement in early and late stages is visible. Shaded bands at late times show the 1σ-scatter due to different [Fe/H].

**Table 2:** Key parameters for the planetary population synthesis model

| | |
|---|---|
| Gas disk viscosity parameter | $2 \times 10^{-3}$ |
| Slope of the initial gas disk | $-0.9$ |
| Exponential cut-off radius of the gas disk | $\propto M_{disk}^{0.625}$ |
| Slope of the initial planetesimal disk | $-1.5$ |
| Planetesimal radius | 300 m |
| Initial number of protoplanets | 50 (100 for 1 $M_\odot$) |
| Initial mass of protoplanets | 0.01 $M_\oplus$ |
| Opacity reduction in the planetary envelope | 0.003 |
| N-body integration time | 20 Myr |

## 3. Results and Discussion

### 3.1 Angular momenta from observational data

We first analyze the distribution of angular momenta of the exoplanet systems from observed data and compare them with our own Solar System. Figure 4 shows scatter plots of orbital (left-panel) and total (mid-panel) angular momenta of confirmed exoplanetary systems (x-axis) versus the ratio of spin angular momentum of the host stars divided by the total orbital angular momentum of all the planets (y-axis). A red-curve across the figure moves from the top to the lower portion illustrating the path of a set of virtual planetary systems with the Sun as the host star, but containing 1 planet, or 2, or 3, … planets at the same masses, sizes and orbits of our Solar System planets: Mercury (M), Venus (V), Earth (E), Mars (M), Jupiter (J), Saturn (S), Uranus (U), and Neptune (N). The positions of such virtual systems are indicated by the black dots on the red-curve. For reference, the positions of angular momenta are related to Earth, Jupiter, and Neptune. An imaginary position of angular momentum relats to Jupiter as if it were in the location of Mercury, is also shown in the left- and mid-panels. It is interesting to see that if Jupiter were in Mercury's orbit, its angular momentum position would be like that of an exoplanet.

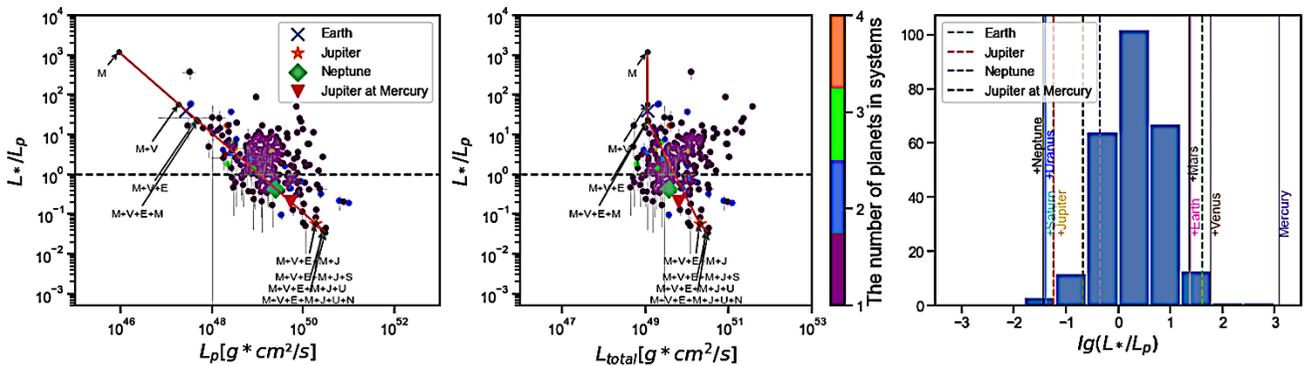

**Figure 4:** A scatter plot of orbital (**left-panel**) and total (**mid-panel**) angular momenta of exoplanetary systems versus the ratio of spin angular momentum divided by orbital angular momentum. A red-curve across the figure moving from top to lower illustrates the path of a set of the virtual planetary systems discussed in the text. The angular momenta related to the Earth, Jupiter and Neptune, as well as the imaginary position of angular momentum related to Jupiter as if it were in the location of Mercury, are also plotted for reference. The **right panel** is the histogram summarizing the number of discovered exoplanet systems in each bin of spin/orbital angular momentum. It is clear that the majority of exoplanets discovered thus far do not have the angular momentum distribution similar to our Solar System planets.

The original idea of Figure 4 (left-panel and mid-panel) is that if the distribution of angular momentum in our Solar System is typical, then we could produce a "prediction curve" of what the angular momentum distribution might look like if more outer exoplanets further away from the host stars would be discovered in the future. Interestingly, there is a gap: the position of angular momenta in such "solar" systems is at either the high-end or low-end compared to the angular momenta computed with the population of currently discovered exoplanets. This is further highlighted by the histogram in the right-panel of Figure 4, in which the x-axis is the ratio of spin angular momenta of stars divided by the orbital angular momenta of the planets and the y-axis is the total count of confirmed exoplanet systems with available data.

It is clear that the majority of exoplanets discovered so far do not have the angular momentum distribution similar to our Solar System planets. In the Solar System, planets within Mars' orbit and closer, are small terrestrial planets, each with a radius of thousands of kilometers, and have a rocky surface. The planets outside Mars' orbit are gas giant planets, with masses ranging from Neptune's $15 M_\oplus$ to Jupiter's $318 M_\oplus$. The gap between the angular momentum distribution for our Solar System in Figure 4 reflects the large separation between the small terrestrial inner planets and gaseous outer giants. Many exoplanets and Solar System planets have obvious differences. For example, Hot-Jupiters have been found with a mass and volume similar to Jupiter; they are giant planets but with orbital distances less than that of Mercury, and their revolution periods are as short as days and hours. Such objects do not exist in the Solar System.

### 3.2 Angular momenta from model simulations and comparison with observations

We now use a population of simulated planets to re-examine angular momenta distributions found in the observed samples. To "filter" the simulated data with observational bias, only the simulated exoplanets with a transit depth > 0.0004, orbital period P > 2 years and with $K_1$ > 10 m/s are counted as the detectable exoplanets.

Angular momentum is a function of the star age. Figure 5 shows that when time goes by, the rotational angular momentum of stars decreases continuously following $10^{-0.02 \cdot \frac{t}{Gyrs}} g \cdot cm^2 \cdot s^{-1}$ based on least-squares fitting (Figure 5 left-panel). Also, to note is that our Sun has very small spin angular momentum compared to most stars. The planet's orbital angular momentum also declines with time but at higher rate of $10^{-0.04 \cdot \frac{t}{Gyrs}} g \cdot cm^2 \cdot s^{-1}$ (Figure 5 mid-panel). The total angular momentum of the system (Figure 5 right-panel), which is the sum of planetary orbital angular momentum and stellar rotation angular momentum, decreases at a rate of $10^{-0.02 \cdot \frac{t}{Gyrs}} g \cdot cm^2 \cdot s^{-1}$. Compared to most exoplanetary systems, both the total orbital angular momentum of the planets and the total angular momentum of our Solar System are among the largest and most similar to the simulations.

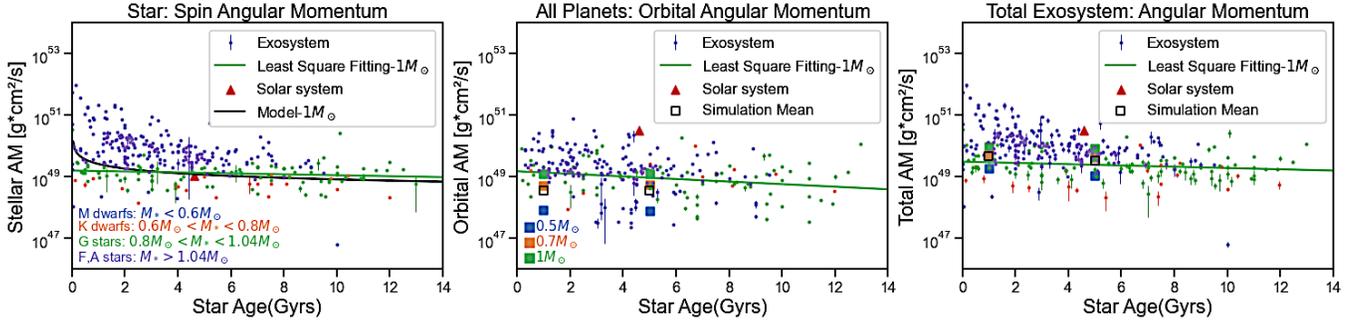

**Figure 5:** Scatter plots of the age of the host star versus spin angular momentum of the host stars **(left-panel)**, orbital angular momentum of individual exoplanets **(mid-panel)**, and total angular momentum of the exoplanetary systems **(right-panel)**. The dots show the data within different stellar mass ranges following the color code shown in the lower left corner in left-panel. The modelled stellar spin evolution follows equation 9 and is shown for different stellar masses. Over Gyr timescales the initial rotation rate of the star no longer matters. Given the good match of the model with the data from different star forming regions by *Gallet & Bouvier* [2015], see Figure 3, the underestimation of exoplanet host spins by the model might point towards an observational bias disfavoring slow rotators. In the right and middle panels, the mean orbital angular momenta from the population syntheses from *Emsenhuber et al.* [2021b] and *Burn et al.* [2021] with applied observational biases (transit depth > 0.0004 and $K_1$ > 10 m/s ) are shown; the mean values for different stellar masses are shown, following the color code in the lower left corner in mid-panel The models do not predict a strong time dependency of orbital angular momenta over Gyr timescales.

Figure 5 shows that the time dependency of the planetary angular momentum is not reproduced by the population synthesis model. While the model does include the removal of planets due to tides, as well as atmospheric mass loss, those processes are not efficient at removing angular momentum. This is because they act on low-angular momentum planets (tides, following *Benitez-Llambay et al.* [2011]) or remove relatively little mass (photoevaporation of atmospheres, *Jin et al.* [2014]). Potential effects on the angular momentum of observable exoplanets that are not included in the model are dynamical self-interactions of the planets after 20 Myr as well as close encounters with other stars. Both of these should be explored in future work to understand the slightly decreasing trend of planetary angular momentum with time. In particular, encounters were shown to be efficient in removing ~10% of the outer Solar System planets (*Stock et al.* [2020]).

Figure 6 and Figure 7 show scatter plots of orbital angular momenta (Figure 6) and total angular momenta (Figure 7) of exoplanetary systems versus the ratio of spin angular momentum divided by orbital angular momentum, respectively. A red-curve across the figure moving from top to lower illustrates the path of a set of virtual planetary systems discussed in the text. The purple dots are the observed data, other colored dots are the simulated data with different host star masses. The simulation results at different time stages (0yr, 100Kyr, 1Myr, 100Myr, 1Gyr and 5Gyr) are shown, respectively. For both Figure 6 and 7, only those simulated data filtered by the observational biases are shown.

After the initial stage (0yr), the number of observable, simulated planets in the protoplanetary disk starts to grow. In both Figure 6 and 7, we see that the angular momentum distribution for the planets form at each time stage appears to line-up along the red-curve and also grouped according to the mass of the host-star.

During the stages embedded in the protoplanetary disks (0yr–1Myr), orbital angular momentum can still increase due to growth of protoplanets. In the second column (100 Myr to 5 Gyr), stellar evolution described in Section 2.3 is found to be the dominating factor. Although the Bern model includes evolutionary processes for planets, it does not significantly alter the angular momentum (see discussion above).

In contrast, the stars converge in terms of their spin angular momentum and reach the *Skumanich* [1972] relation after their pre-main sequence phase. The scatter due to different metallicities is considered here (see Section 2.3) but estimated to be of a small magnitude following the modern breaking law cases [*Matt et al.* 2015, *van Saders & Pinsonneault* 2013] investigated by *Amard et al.* [2019]. In Figure 4, we see that in our sample the sun is a relatively slow rotator. This is not recovered in Figure 2, comparing the sun to open clusters where the sun is found to be average. Therefore, there might be a significant selection effect in the exoplanetary data used here. Indeed, it is easier to measure fast rotation rates due to the decreased observational time needed.

A prominent feature in the angular momentum distribution of synthetic planets is the "planetary desert" [*Ida & Lin* 2004]. It separates systems where no planet was reaching critical core masses to trigger runaway gas accretion [e.g., *Pollack et al.* 1996] with those where at least one planet was able to accrete a significant amount of gas. Given the presence of two orders of magnitude more gas than solids, planets able to accrete gas have the potential to dominate the orbital angular momentum of the system. In Figure 5, the planetary desert is located at angular momenta below Jupiter's corresponding to $10^{50}$ *g·cm²/s*. *Burn et al.* [2021] found that giant planets form around stars with masses of 0.5 $M_\odot$ or larger, which is why the large orbital angular momentum systems are absent in Figures 5 and 6 for stellar masses of 0.1 and 0.3 $M_\odot$.

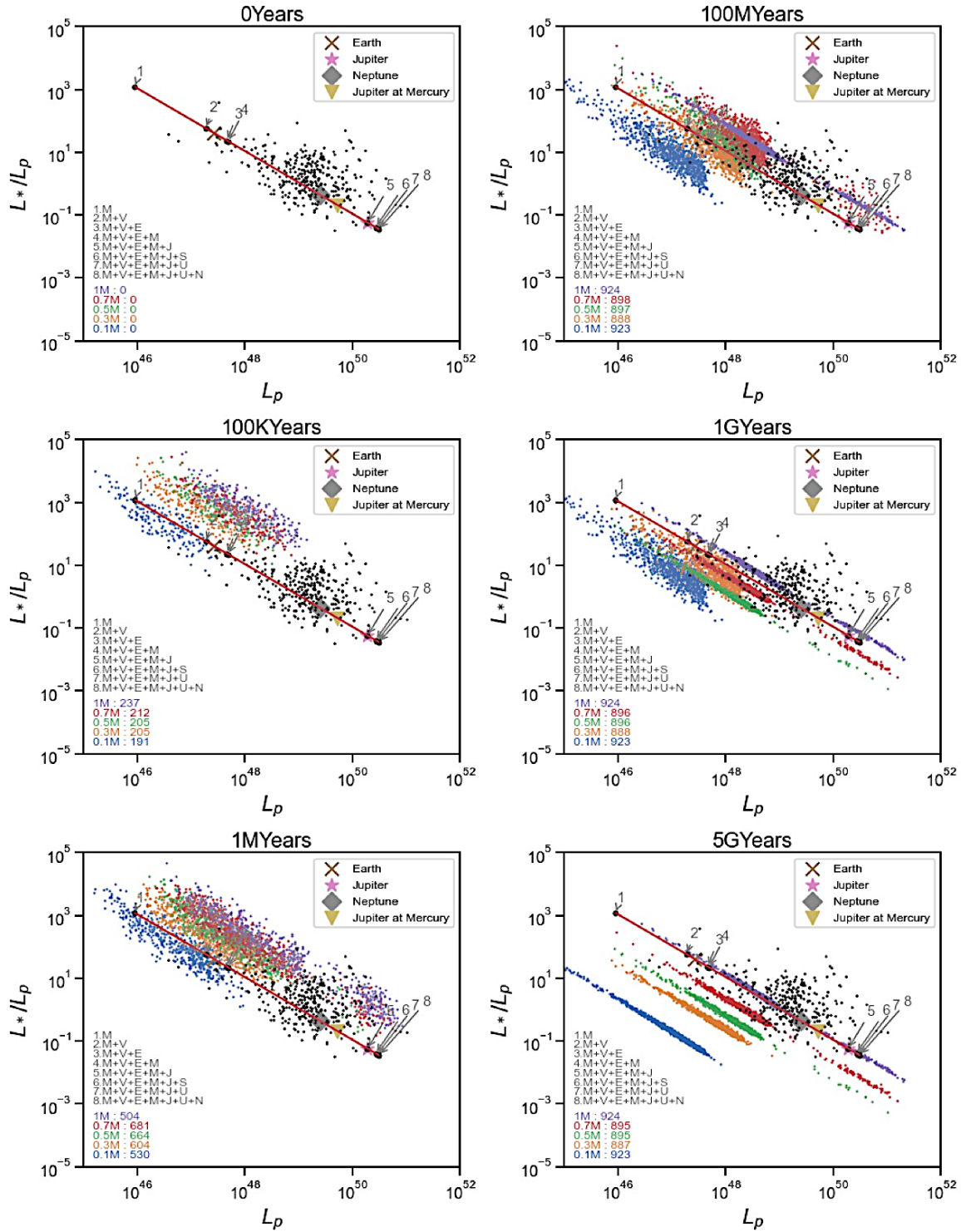

**Figure 6:** Scatter plots of orbital angular momenta of exoplanetary systems versus the ratio of spin angular momentum divided by orbital angular momentum. A red-curve across the figure moving from top to lower illustrates the path of a set of the virtual planetary systems discussed in the text. The black dots are the observed data, other colored dots are the simulated data with different host star masses. The numbers at the lower left corner show the number of detectable systems around the star with different mass showing with the same color code. The simulations results are outputted at different time stages (0 yr, 100Kyr, 1Myr, 100Myr, 1Byr and 5Byr) respectively.

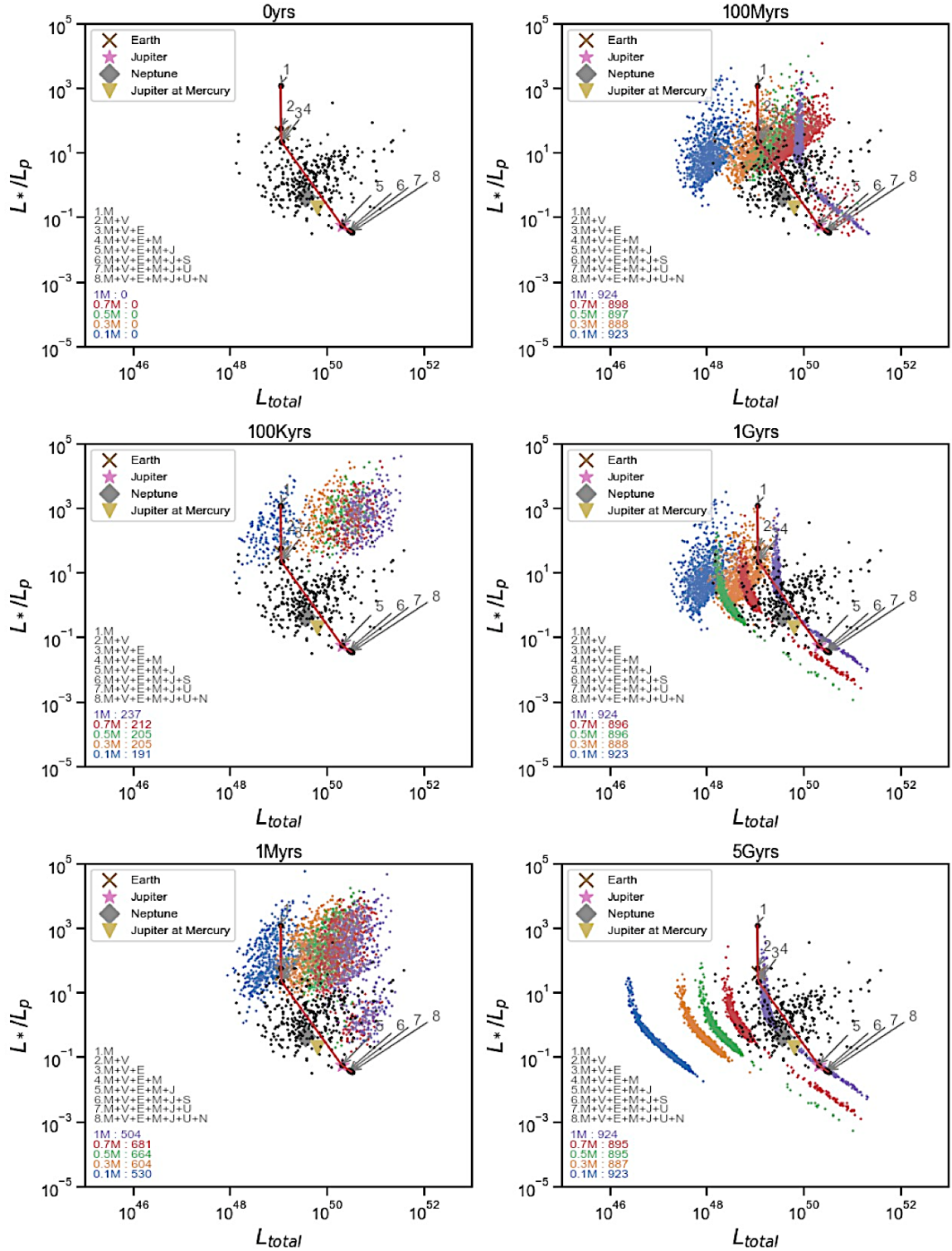

**Figure 7:** Scatter plots of total angular momenta of exoplanetary systems versus the ratio of spin angular momentum divided by orbital angular momentum. A red-curve across the figure moving from top to lower illustrates the path of a set of virtual planetary systems discussed in the text. The black dots are the observed data, other colored dots are the simulated data with different host star masses. The numbers at the lower left corner show the number of detectable systems around the star with different masses being shown with the same color code. The simulation results are output at different time stages (0yr, 100Kyr, 1Myr, 100Myr, 1Gyr and 5Gyr) respectively.

However, this feature is not discernible in the observed distribution of angular momenta. Indeed, this agrees with the finding of *Suzuki et al.* [2018] based on microlensing surveys; they did not find a drop in planetary occurrences at the theoretically predicted location of the planetary desert (or sub-Saturn-mass desert). For transit [*Thompson et al.* 2018] and radial velocity [*Mayor et al.* 2011] datasets the interpretation is less conclusive, but the desert cannot be confirmed [*Bennett et al.* 2021]. Numerical simulations also show that the 1D accretion treatment used in the Bern model [*Mordasini et al.* 2012] is not able to accurately capture the three-dimensional accretion process in all mass regimes [*Szulagyi et al.* 2014, *Moldenhauer et al.* 2021]. This might be especially relevant if a circumplanetary disk forms instead of an envelope [*Szulagyi et al.* 2016].

Therefore, we retrieve here in a different parameter space – orbital angular momentum instead of mass – the discrepancy of 1D theoretical accretion and observations. Our analysis is not yielding more rigorous constraints on the issue since we take a simple approach to modeling observational biases and we ignore sampling effects.

## 4. Summary

The angular momenta of planetary systems are examined using the data from NASA Exoplanet Archive as well as by a model simulation using planetary population synthesis. We found that a majority of exoplanets discovered so far do not have the angular momenta distribution similar to our Solar Systems planets. The total orbital angular momenta of the exoplanets are considerably smaller than that of the Solar System planets.

When the host star ages over time, the total angular momentum of the exoplanetary system decreases at a rate about $10^{-0.02 \cdot \frac{t}{Gyrs}} g \cdot cm^2$, which contributed from declines of angular momenta both from the spin of the star $10^{-0.04 \cdot \frac{t}{Gyrs}} g \cdot cm^2$) and the orbiting of the exoplanets ($10^{-0.02 \cdot \frac{t}{Gyrs}} g \cdot cm^2$). The decrease of stellar angular momentum is consistent with the model simulation; however, time-dependency of orbital angular momentum of the exoplanets is not reproduced by the models.

The simulated planetary angular momentum distributions at each evolutionary stage appear to line-up and was able to be grouped according to the mass of the host star. Moreover, the "planetary desert" feature in the synthetic planets is not perceptible in the observed distribution of angular momenta. These comparisons between observation and model simulation in angular momentum for exoplanetary systems provide significant insights on the gaps in both data and understanding that need future improvement in this field.

**Acknowledgments**

This research was supported by the Jet Propulsion Laboratory, California Institute of Technology, under the contract with NASA. We acknowledge the funding support from the NASA Exoplanet Research Program NNH18ZDA001N. R.B. acknowledges the financial support from the SNSF under grant P2BEP2_195285.

**Data Statement**

The data underlying this article are available in the article and in its online supplementary material. For additional questions regarding the data sharing, please contact the corresponding author at Jonathan.H.Jiang@jpl.nasa.gov.